# South Atlantic continental margins of Africa:
# a comparison of the tectonic vs climate interplay on the evolution of equatorial west Africa and SW Africa margins

Michel Séranne, Zahie Anka
UMR 5573 Dynamique de la Lithosphère
CNRS / Université Montpellier 2
34095 Montpellier cedex 05

*corresponding author:*
Michel Séranne
seranne@dstu.univ-montp2.fr
+(33) (0)467 14 67 42 61 / +(33) (0)467 54 48 11

**Abstract**

Africa displays a variety of continental margin structures, tectonic styles and sedimentary records. The comparative review of two representative segments: the equatorial western Africa and the SW Africa margins, helps in analysing the main controlling factors on the development of these margins. Early Cretaceous active rifting south of the Walvis Ridge resulted in the formation of the SW Africa volcanic margin that displays thick and wide intermediate igneous crust, adjacent to a thick unstretched continental crust. The non-volcanic mode of rifting north of the Walvis ridge, led to the formation of the equatorial western Africa margin, characterised by a wide zone of crustal stretching and thinning, and thick, extensive, synrift basins. Contrasting lithologies of the early post-rift (salt vs shale) determined the style of gravitational deformation, whilst periods of activity of the decollements were controlled by sedimentation rates. Regressive erosion across the prominent shoulder uplift of SW Africa accounts for high clastic sedimentation rate during Late Cretaceous to Eocene, while dominant carbonate production on equatorial western Africa shelf suggests very little erosion of a low hinterland. The early Oligocene long-term climate change had contrasted response in both margins. Emplacement of the voluminous terrigenous Congo deep-sea fan reflects increased erosion in equatorial Africa, under the influence of wet climate, whereas establishment of an arid climate over SW Africa induced a drastic decrease of denudation rate, and thus reduced sedimentation on the margin. Neogene emplacement of the African superswell beneath southern Africa was responsible for renewed onshore uplift on both margins, but it accelerated erosion only in the Congo catchment, due to wetter climatic conditions. Neogene high sedimentation rate reactivated gravitational tectonics that had remained quiescent since late Cretaceous.

# 1. Introduction

The South Atlantic continental passive margins of Africa comprise the major depocentres on the African plate. They reach 15 km in thickness off Angola and extend hundreds of kilometres across the continent-ocean transition. Sediment accumulation not only records the evolution from the continental rifting and break-up of Gondwana during latest Jurassic - Early Cretaceous, but also provides a complete record of the post-rift evolution of the margin and of adjacent continental Africa.

Following hydrocarbon discoveries in the second half of the 20th century, there has been an increasing amount of literature published on these margins. The rate of publication increases with both the release of industrial data, and the extension of the exploration acreage toward deeper bathymetry (Cameron et al., 1999; Mohriak and Talwani, 2000; Arthur et al., 2003).

Integrated studies of continental margins reveal the complex interactions that exists between internal and external forcings on their formation and development (e.g. Séranne, 1999; Lavier et al., 2001; Lucazeau et al., 2003). Sedimentary depocentres of continental margins are currently analysed with the concepts of sequence stratigraphy, but sediment flux is pivotal in developing predictive stratigraphy. It thus appears that continental margin studies must also integrate the study of the river drainage on adjacent continents, in order to understand and quantify the terrigenous flux. In this respect, the South Atlantic margin of Africa covers a range of climatic zones and consequently varying qualities and quantities of terrigenous imput, where the climatic control on margin's stratigraphy can be tested.

The aim of this paper is to review and compare the major features of two representative segments of these margins. An exhaustive review of both margins being beyond the format of this contribution, we have limited the comparison to selected features, with emphasis on the driving mechanisms.

This contribution is based on data and interpretation made available to academia through publications. We are well aware that an extensive amount of additional, although proprietary and confidential, data exists. Progressive release of such data will probably confirm or modify some of our interpretation.

# 2. Structural framework

The area reviewed here extends from equatorial Africa to the southern tip of the continent (**Fig. 1**). To the north, it is separated from the equatorial transform margin and associated Niger delta by the Cameroon Volcanic Line. The southern boundary is placed along the Falkland-Agulhas Fracture Zone.

South Atlantic results from rifting of the Gondwana during Late Jurassic to Neocomian. Break-up and oceanic accretion from late Neocomian (in the south) to Aptian (in the North) (e.g. Uchupi, 1989) marks the beginning of drifting of South America and Africa.

Kinematics of the South Atlantic opening is still controversial (see review in Moulin, 2003). However, the following statements are fairly well established: 1) some intraplate deformation is observed on the continents, 2) pre-rift fits of rigid plates of Africa and South America leave significant gaps between margin hinge lines, and 3) drifting of the southern segment of the South Atlantic predated opening of the central segment. Kinematics of the South Atlantic and its continental margins are strongly dependent upon the understanding of the relationships between rifting and the Tristan da Cunha hot spot. The latter was responsible for the formation of the prominent morpho-tectonic feature of the South Atlantic (Rio Grande Rise and Walvis ridge (O'Connor and Duncan, 1990), and emplacement of basaltic plateaux of Parana in South America and Etendeka in Namibia (Renne et al., 1996).

The South Atlantic margin of Africa is divided into two major segments: the margins of equatorial western Africa and SW Africa, separated by the Walvis Ridge. Each one has been subdivided into many coastal sedimentary basins, according to transform faults or basement highs segmentation and/or according to political and economic boundaries; this results in a complex and not always consistent sedimentary basins nomenclature. In addition, the equatorial western Africa margin and the SW Africa margin are facing the Angola Oceanic Basin and the Cape Oceanic Basin, respectively, and significant sedimentary depocentres rest on oceanic crust.

The total thickness of sediment is unevenly distributed along the South Atlantic margins of Africa. **Figure 1** represents the depocentres along the coastline, recognised by seismic reflection surveys (Emery et al., 1975a; Emery et al., 1975b; Moulin, 2003; Anka, 2004). The equatorial western Africa margin includes one of the thickest depocentres ,which extend partly onshore, whereas the smaller depocentres of the SW Africa margin are restricted to the offshore domain. This map clearly shows the close relationship between the main depocentres and the river drainage supplying terrigenous flux to the margin. This is particularly well expressed for the Ogooue, Congo, and Orange rivers mouths where the largest depocentres are found. However, river drainage - depocentre relationships are not straightforward and need to be refined: some significant depocentres are not related to rivers (e.g. Luderitz Basin, Namibia); a small river (Ogooue) fed a depocentre as big as that of a giant drainage (Congo, the 2nd World's largest catchment). This can be accounted for by the history and distribution of subsidence, combined with varying erosion processes and localisation, through syn-rift and post-rift evolution of the margin.

## 3. Lithospheric structure and rifting

### 3.1. The South-West Africa volcanic margin

*Crustal structure*
Insight into the deep structure of the SW African continental margin relies on multichannel reflection and refraction seismic surveys available for Namibia (Gladczenko et al., 1997; Gladczenko et al., 1998; Bauer et al., 2000). They display specific features of volcanic margin **(Fig. 2b)**. The Namibia margin is therefore used as the representative segment of the SW African margin, south of the Walvis Ridge.

According to these studies, continental crust thickness decreases moderately: from over 45 km inland, to 25 km (from Moho to top basement) at about 100km west of the coastline. A very thick (+/-25 km) intermediate igneous crust extends over 200 km width. It comprises a lower lenticular 7 km-thick body, of anomalously high-velocity, interpreted as underplated material, a 8-12km thick interval, corresponding to a complex of mafic intrusions. Finally, reflections profiles display in the upper interval seaward dipping reflectors (SDR) that represent subaerial volcanic flows emplaced during rifting. Westward transition to normal oceanic crust occurs prior to the M4 magnetic anomaly (Bauer et al., 2000), which represents the oldest (127Ma, Hauterivian-Barremian transition) evidence of mid-oceanic ridge activity on this transect. The upper section of the rifted continental crust displays small (< 5 km thick) fault-bounded, synrift basins that pass westward to the SDR's. There is evidence of an ancient, well developed, rift basin unconformably overlain by the Mesozoic structures (Gladczenko et al., 1998), interpreted as Permian to Triassic in age (Karoo Rift, Clemson et al., 1999). Post-rift sedimentary wedge extends mostly west of the continent-ocean boundary (COB), reaching a maximum thickness (circa 6 km) above the transitional igneous crust. The crustal section **(Fig. 2b)** shows a lateral offset of syn- and postrift depocentres: the maximum post-rift thickness (4.5 km) is found above the transitional igneous crust, some 100 km west of the main synrift depocentre. This setting contrasts with the usual thermally-driven subsidence of stretched lithosphere, where maximum post-rift subsidence is centred onto synrift basins. It may reflect the interference of a load-driven subsidence introduced by the emplacement of thick and dense intermediate igneous crust in the continental margin.

*Rifting*
Intracontinental rifting leading to the opening of the future South Atlantic occurred during late Jurassic - early Cretaceous. It was accommodated to the south by right-lateral motion on the Agulhas-Falkland Fracture Zone. However, this rifting was preceded by a "Basin and Range type", late-orogenic extension of a Pan-African tectonically thickened crust (Light et al., 1993). This phase allowed deposition of the extensive Permian - Jurassic Karoo continental formation, and development of a shallow marine embayment in the most subsiding axis, localised between the two future continents (Light et al., 1993). Clemson and others (1997; 1999) recognized an early Triassic syn-rift and and late Triassic-Middle Jurassic post-rift sequences, bounded by reactivated Damara (Pan-African) fold belt structures, which, later controlled the segmentation of the South Atlantic rifting.

Fault-bounded sequences related to South Atlantic rifting are divided into syn-rift I and syn-rift II in all SW African basins: Orange Basin (Gerrard and Smith, 1982; Jungslager, 1999), Luderitz and Walvis Basins (Maslanyj et al., 1992; Light et al., 1993), and they are separated by an angular unconformity ("R") of Valanginian age (Gerrard and Smith, 1982). The age of rift onset (marked by the "T" unconformity of (Gerrard and Smith, 1982) is correlated with late Jurassic. Synrift sequences are characterised by continental "red-beds" and volcanics rocks, with rare lacustrine shale occurrences (Jungslager, 1999). Syn-rift volcanism occurred both in the basins and as continental flood basalts (e.g. Etendeka province in northern Namibia dated to 132 Ma (Renne et al., 1996). In Namibia, there is compelling seismic evidence for lava flows, sills and dykes interbeded with, and intruding the synrift sediments, as well as borehole data drilling hundreds of meters of basalt successions (Jackson et al., 2000). Such volcano-clastic sequences pass basinward and upward to seaward-dipping reflectors (SDR) (Gladczenko et al., 1997). They were emplaced, at surface level, during the initial stages of continental lithosphere separation. Existence of two (inner and outer) wedges of SDR's across a 200 km-wide zone **(Fig. 2b)** (Bauer et al., 2000) suggests a complex evolution over a significant time interval before complete break-up. Normal thickness oceanic crust, related to the activity of a spreading ridge, extends seaward of the M4 magnetic anomaly, suggesting that drifting started at around 127 Ma.

### 3.2. The non-volcanic equatorial western Africa margin

*Crustal structure*
The margin structure north of the Walvis Ridge has been investigated by several deep-seismic experiments since the early 90's and an increasing amount of data is being released and reprocessed (Rosendahl et al., 1991; Wannesson et al., 1991; Meyers and Rosendahl, 1991; Moulin et al., 2001; Dupré, 2003; Contrucci et al., 2004). The continental margin across the Angola segment comprises a 150 km-wide zone of extremely thinned continental crust (5 to 8 km thick), which is overlain by a 10-12 km sedimentary prism, and underlain by an anomalously high velocity zone, at the transition with moderately extended continental crust. The origin of this anomalous interval could represent either an abnormal lower crust or mantle (Contrucci et al., 2004), or underplated material (Dupré, 2003). The narrow transition between extended continental crust and normal oceanic crust occurs a few kilometres landward of the Angola Escarpment (a prominent north-south bathymetric step at the slope-basin boundary). The crustal structure and the seismic imaging of the sediment-basement interface markedly differs from those of the Namibia volcanic margin: there is no evidence for a thick transitional igneous crust, and the possible SDR's on the Angola margin postulated by (Jackson et al., 2000) are contradicted by more recent seismic data (Contrucci et al., 2004; Moulin, 2003). In spite of the scarce occurrence of basaltic lavas related to the Etendeka province (Marzoli et al., 1999), the crustal structure of the Angola margin and the < 10 km thick oceanic crust characterise a non-volcanic margin (Contrucci et al., 2004).

*Rifting*
Landward of the "Atlantic Hinge Line" (Karner et al., 1997), and in the northern part of the Congo Basin, the rifting-related structures comprise several extensional faults and tilted syntectonic basinfills (Teisserenc and Villemin, 1989; Meyers et al., 1996). However, these typical rift-related structures are markedly absent in the middle and lower margin of equatorial western Africa margin (Moulin, 2003; Contrucci et al., 2004). Across this segment **(Fig.2a)**, the continental crust rapidly thins down to less than 10 km over a distance of 50km. Seaward of the tilted and faulted blocks, extends a 2-to-5 km-thick and 150km-wide sequence of parallel basin-fill. To the east, the pre-salt strata onlap a pre-rift high, and abut against a basement high close to the continental-oceanic crust transition.

Extensional block faulting occurred landward of the "Atlantic Hinge zone", during Berriasian to early Barremian (Teisserenc and Villemin, 1989), and was responsible for the deposition of 2 sequences (rift 1 an rift 2) of lacustrine black shale, and turbidites, followed by a shoaling up sequence of prograding deltaic / fluvial system (Robert and Yapaudjian, 1990). These two syn-rift sequences are overlain by a third synrift sequence of Barremian to early Aptian shallow lacustrine shales and sandstones. The syntectonic feature of this last sequence is displayed in the onshore basin (i.e. in Gabon, Teisserenc and Villemin, 1989). Seismic correlation across the pre-rift basement high, with the western aggrading depocentre, is still questioned **(Fig. 3)**. This unfaulted, 5km-thick package is either contemporaneous with the synrift 1,2 and 3 sequences, either a time equivalent to the synrift 3 only. It is suggested that accommodation seaward of the Atlantic Hinge Zone was delayed with respect to the block faulting (Moulin, 2003), and continued during the last phase of rifting that affected the entire margin, during late Barremian to early Aptian.
These observations indicate that the Gabon-Congo-Angola margin was thinned by extensional faulting landward of the Atlantic hinge zone, during Early Cretaceous (the oldest synrift sequence is not known in Angola, but it predates Barremian). During this interval, homogeneous subsidence of the western part of the margin, mostly during the Barremian, suggests lower-crustal thinning process (Karner et al., 2003). Asymmetric extension across a low-angle crustal-scale normal fault may account for the observations.

Continental break-up post-dates this last rifting event. To the east, the tilted blocks are unconformably overlain by the transitional sequence comprising evaporites, deformed by halokinesis. To the west, seaward of the basement high, the wide pre-salt depocentre is parallel to the base salt.

Based exclusively on published data and interpretations, this complex syn-rift sequence seems to indicate a multi-stage rifting process, that reflects the diachronous opening of the future South Atlantic (Davison and Bate, 2004).

## 4. Post-rift stratigraphic evolution

### 4.1. Rift-Drift transition

It is widely accepted that rift-drift transition on the South Atlantic margins becomes younger toward the north. The end of rifting can be seismically characterised with break-up unconformity that seals the extensional faults, and with packages of sub-parallel reflectors that onlap the basement (Braun and Beaumont, 1989). The beginning of the drift sequence may be correlated with the onset of oceanic spreading ridge; drifting is therefore as old as the oldest magnetic anomaly (Austin and Uchuppi, 1982).

The onset of the drift sequence on the SW Africa margin, south of the Walvis, was associated with the "AII" (Atlantis II) unconformity (Emery et al., 1975a), dated in DSDP hole 361, to the Aptian-Albian boundary (Bolli et al., 1978). However, synrift sequences and normal faults are sealed by Barremian packages of sub-parallel, conformable reflectors visible beneath AII and above diverging, fault-bounded, synrift reflectors (Austin and Uchuppi, 1982). This observation lead (Jungslager, 1999) and (Brown et al., 1995) to place the "break-up unconformity", equivalent to the drift onset, at the base of the Barremian. In addition, identification of magnetic anomalies M11 and G along the SW African Margin suggested to Rabinowitz and LaBreque (1979) and Austin and Uchuppi (1982) that oceanic accretion and therefore continental break-up occurred as early as Valanginian. However, later interpretations on the Namibia margin correlate pre-M4 magnetic anomalies to SDR's (Bauer et al., 2000), which are interpreted as an initial, intermediate, form of oceanic crust emplaced from a subaerial (above sea-level) spreading centre (Talwani and Abreu, 2000). These authors consider that SDR's post-date terrigenous synrift sedimentation and the 130 My-old Etendeka continental flood basalt (Renne et al., 1996), which leads to a late Hauterivian to early Barremian age for the age of initial break-up. The rift-drift transition therefore starts above the Barremian unconformity. In the inner part of the margin, it corresponds to a sag basin characterised by restricted marine environments, followed by a major flooding event during early Aptian time. The latter was associated with deposition of anoxic black shale. From late Aptian onward, a siliciclastic ramp prograded westward (Brown et al., 1995), while the base of the prograding sequences correlates basinward with the AII unconformity.

On the equatorial western Africa margin, the transitional rift-drift interval unconformably rests on the continental Barremian synrift sequences. The unconformity, interpreted as the break-up unconformity, is associated with a short hiatus of the lowermost part of the Aptian (Logar et al., 1983). The interval starts with a transgresssive clastic sequence grading upward from fluvial sandstones and lagoonal shales to thick layers of evaporites (Teisserenc and Villemin, 1989). Deposition of evaporites in the "Salt Basin" (Belmonte et al., 1965) makes the distinctive feature of equatorial western Africa rift-drift transitional sequence. Deposition of salt seems diachronous: early salt deposition locally occurred during synrift in Congo (Karner and Driscoll, 1999) and in Kwanza (Davison and Bate, 2004); however, generalised post-rift salt deposition occurred during middle to late Aptian. The confined basin was separated from the World ocean by the Walvis Ridge, to the south, and the still connected Amazon and West African Cratons, to the north (Davison and Bate, 2004). The Salt Basin results from precipitation of evaporites (mostly massive halite, topped by anhydrite from saturated brines (Teisserenc and Villemin, 1989). Sea-water was fed over sills to the evaporitic basin in several cycles, thus accounting for the 700-1500m thickness of salt (Stark et al., 1991). Interestingly, Tethyan ostracods found in Barremian syn-rift sequences suggests that the earliest connection with the World Ocean was achieved across the north. Aptian evaporites have been intensely deformed and transferred down slope by later salt tectonics (see below), and it is thus difficult to restore the original thickness. The Salt Basin correlates with the "Aptian Anoxic Basin" of the SW Africa margin (Jungslager, 1999), the drastically different paleogeography resulting from the Walvis Ridge that sectioned off part of the South Atlantic.

**4.2. Drift sequence**

On the SW Africa Margin, after the major flooding event in the early-Aptian, an aggrading marine sequence was deposited over 500 km width, which characterises an even subsidence across the margin, from the stretched continental crust to the newly accreted oceanic crust **(Fig. 4)**. From Albian to Turonian, an exclusively siliciclastic shelf prograded over100 km west of the hinge line (Brown et al., 1995) as a result of massive and continuous terrigenous sediment input (Rust and Summerfield, 1990). High rates of early post-rift, thermal subsidence were balanced by sedimentation on the shelf. Senonian sediments prograded across the shelf and were deposited on the slope and continental rise (Emery et al., 1975a, Dingle and Robson, 1992) while seismic stratigraphy suggests erosion on the innermost shelf (Muntingh and Brown, 1993). This indicates that continued high rates of clastic sedimentation (Rust and Summerfield, 1990) were exceeding the declining thermal subsidence rates on the shelf. Large volumes of terrigenous sediments were derived from the small coastal watersheds of Namibia and the Orange drainage, where up to 3 km denudation was achieved during late Cretaceous (Gallagher and Brown, 1999). During Tertiary, sediments by-passed the shelf and were deposited in a thin wedge, mostly on the slope.

The post-rift stratigraphic architecture of equatorial western Africa differs from that of the SW Africa margin in several ways **(Fig. 4)**. The marine transgression started with the deposition of evaporites and followed with the deposition of shallow carbonate (dolomites then limestone) during Albian (Pinda or Madiela Fm.). These aggrading carbonates are markedly absent in SW Africa margin, and extended over a very wide area (>200 km) as shown by the DSDP hole 364 (Bolli et al., 1978). In spite of high subsidence rate of the early thermal recovery of the stretched lithosphere, carbonate production kept-up with accommodation. The width of the aggrading shelf/ramp indicates an unusual mode of subsidence with little or no differential subsidence across the entire margin (Séranne, 1992; Séranne et al., 1992). Seaward of this carbonate shelf/ramp, that was affected by early halokinesis, probably extended a deep-sea fan fed by a source close to the present-day Congo, as revealed by seismic reflection surveys (Anka, 2004). From late Turonian onwards, the depositional profile drastically changed. Carbonates did not keep up anymore with accommodation and terrigenous sedimentation progressively took over. As a result, bathymetry increased offshore and a ramp profile developed. Except on the landward side of the margin, the rate of sedimentation dropped drastically during Senonian and early Paleogene **(Fig. 4)**, although basinal sedimentation remained significant during Senonian in Gabon (Teisserenc and Villemin, 1989).

**4.3. Late drift : Oligocene-Present deep-sea fans**

A major stratigraphic reorganisation of the equatorial western Africa margin occurred in the Early Oligocene (Séranne, 1999; Séranne et al., 1992). It was characterised first by a major submarine erosion of the ramp (Séranne et al., 1992; McGinnis et al., 1993), that removed some 500m of stratigraphic section in the outer shelf (Nzé Abeigne, 1993; Nzé Abeigne, 1997; Lavier et al., 2000). Second, the deeply-incised erosional surface was down-lapped by a clastic prograding wedge that extended basinward to a very large turbidite deep-sea-fan: the Congo fan (Anka and Séranne, 2004). The prograding shelf is mostly by-passed by sediments. Sedimentation on the slope is dominated by fine terrigenous sediments and hemipelagic drape reworked by bottom currents, probably in relation with upwelling (Séranne and Nzé Abeigne, 1999). On the lower slope and deep basins, turbidites accumulated downstream of major rivers (Ogooue (Mougamba, 1999), Congo (Droz et al., 2003). Turbidites are fed to the lower slope and basin by canyons cross-cutting the shelf, within the lower Miocene stratigraphic interval (Wonham et al., 2000). However, the present Congo canyon is the most spectacular example of such erosive structures (Heezen et al., 1964; Babonneau et al., 2002) that confines and transfers turbidite currents to the lower slope and deep-sea fan (Turakiewicz, 2004).

On SW Africa margin, a break in the stratigraphy is also documented at the Eocene-Oligocene transition (Rust and Summerfield, 1990). However, the sedimentation hiatus of the Oligocene ($A^u$ unconformity) is followed by a period of sedimentation accumulation reduced by half in comparison to late Cretaceous values (Rust and Summerfield, 1990). To the south, the Oligocene shelf erosion and canyon-cutting event was associated with turbidite deposition on the slope (Dingle and Robson,

1992). Seismic reflection profiles across the slope display a prograding clastic wedge affected by slumps and mass transport deposits, reworked by bottom currents (Dingle and Robson, 1992; Berger et al., 1998; Bagguley and Prosser, 1999). The onset of a bottom current in the Oligocene was related to reorganisation of ocean circulation, following the opening of the Drake Passage and the isolation of Antarctica (Kennett and Stott, 1990) and was responsible for deep basin erosion (Siesser, 1978; Tucholke and Embley, 1984).

The regional expression of the Oligocene-Present late drift sequence strongly differs on the two reviewed segments **(Fig. 5)**. On the SW Africa margin, the Oligocene-Present interval is less than 1.5km thick, and is evenly distributed along the upper slope, seaward of the late Cretaceous depocentre, between the Walvis Ridge and the Agulhas-Falkland Transform Zone (Rust and Summerfield, 1990). In particular, there is no marked depocentre offshore the Orange river mouth, which suggests that sediments derived from this presently major catchment (953000 $km^2$, (Dingle and Hendey, 1984) did not dominate the other sedimentary sources made of the small coastal catchments.

In contrast, the post-Oligocene sequence of the equatorial western Africa margin is characterised by large depocentres that are localised at the outlet of the main rivers, thus clearly indicating the sedimentary source **(Fig. 5)**. The total volume of that sequence was evaluated to $1.2 \times 10^6$ $km^3$, making up half of the total post-rift sequence (Leturmy et al., 2003). The thickest (> 4 km) and largest (> $0.5\ 10^6$ $km^3$) post-Oligocene depocentre lies across the margins of Congo and Angola, and extends over the oceanic crust to the Congo deep-sea fan. The Congo deep-sea-fan has been prograding since Oligocene (Anka and Séranne, 2004) interacting in a complex way with the evolution of the margin (Anka, 2004). Estimation of Tertiary denudation from relic surfaces in the Congo river catchment and on the smaller coastal river catchments accounts for the bulk of terrigenous sediments accumulated during that period (Leturmy et al., 2003), although increasing rates of accumulations can be documented through the Neogene (Mougamba, 1999; Séranne, 1999; Lavier et al., 2001; Anka, 2004). This reflects an increasing erosion rate on the adjacent continent, during the Neogene.

Many debates have been going on about the origin of the Early Oligocene transition from aggradation to progradation stratigraphic pattern on the South Atlantic margins of Africa. World-wide occurrence of synchronous similar stratigraphic turn-over observed on continental margins points to a climatically-driven, global change, corresponding to the greenhouse-icehouse transition (Séranne, 1999; Lavier et al., 2001). Other possible causes could be related to tectonics that affects the margin. This will be discussed later, in the light of the post-rift tectonic evolution of the margins.

## 5. Post-rift tectonics

### 5.1. Gravitational tectonics

Post-rift sequence of SW Africa margin displays gravitational tectonics in the slope area **(Fig. 4)**. They consist in extensional, seaward-dipping rotational listric growth faults, detached above decollements along which, seaward movement is transferred and balanced by imbricated thrusting in the lower slope **(Fig. 4 and 6)** (Light et al., 1993; Muntingh and Brown, 1993; Brown et al., 1995). In the Orange basin, at least 3 stratigraphic shale units have developed decollement-levels within the drift sequence. The deepest decollement corresponds to shales from a marine-condensed sections of the Cenomanian-Turonian transition (Muntingh and Brown, 1993). An intermediate decollement level occurs in the lower Campanian and the shallowest gravitational system detaches above the base Tertiary decollement (Jungslager, 1999). Listric extensional faults developed only above dipping decollements, i.e. in the paleoslopes; in contrast, the prominent Aptian anoxic shale interval, which is almost horizontal, did not generate a decollement for the overlying sequence. The sedimentation rate **(Fig. 6)** is the other determinant parameter for triggering gravitational tectonics, as shown by the northwards decreases of amplitude and extent of gravitational tectonics, away from the Orange depocentre. Indeed, the thin Tertiary sequence off Namibia, is not affected by gravitational deformation (Light et al., 1993). During deposition of the whole drift sequence, shallow slumping affected the upper slope of southernmost SW Africa margin, and during Neogene slumping also affected the slope off the Orange River mouth (Dingle and Robson, 1992).

The equatorial western Africa margin displays some of the most spectacular salt tectonic structures. Their study has been the topic of numerous works (e.g. Pautot et al., 1973; Duval et al., 1992; Lundin, 1992; Liro and Cohen, 1996; Spathopoulos, 1996; Raillard et al., 1997; Marton et al., 2000;Tari et al., 2003; Untenehr and de Clarens, 2004, Fort et al., 2004). Unlike in SW Africa margin, there is a single decollement level located within the mid-Aptian evaporites that underlie the entire rifted margin of the equatorial western Africa, from onshore basins to the continent-ocean boundary.

Structures are organised across the margin in a similar way (Raillard et al., 1997; Marton et al., 2000; Tari et al., 2003). From the onshore outcrops of the shelf area, late Cretaceous sequences are affected by extensional listric growth faults detached in the salt, with salt rollers rising in the footwall. Extensional offset did not exceed the thickness of the Cretaceous sequence, so the fault blocks are known as "pre-rafts". Due to an initially thin salt layer and to the transfer towards salt rollers, the salt thickness is reduced to the thickness of the decollement. The hinge zone is marked by a major, seaward-facing, extensional listric fault that bounds a distinctive half-graben displaying growth structures. The upper slope in underlain by several extensional listric faults with offset large enough to have broken apart the Cretaceous sequence and separated the fault-blocks with Tertiary depocentres ("rafts"). In the lower slope, series of extensional faults and diapirs, locally associated with evidence of inversion and thrusts, affect the whole post-rift sequence. This association of structures characterises seaward translation of the post-rift package. Finally, the base of slope is dominated by imbricated thrusts, large-scale diapirs and canopies (Cramez and Jackson, 2000 Marton et al., 2000; Gottschalk et al., 2004; Jackson et al., 2004) characterising shortening that balances extension of the upper margin. A contractional front (the "Angola Escarpment") is located at the western end of salt, close to the ocean-continent boundary. Detached sequences may be translated only several kilometres over the oceanic crust (Anka, 2004).

Analysis of the growth structures allows deciphering o the timing of salt tectonics in equatorial western Africa. Early growth structures within the Albian carbonate witness the onset of extensional deformation, that is sealed by Senonian deposits on the shelf (Séranne et al., 1992; Marton et al., 2000; Valle et al., 2001), while extension continues until Campanian on the upper slope (Spathopoulos, 1996; Cramez and Jackson, 2000). A second major phase of extension occurred on the slope during the Oligocene: reactivation of faults resulted in the formation of the rafts (Duval et al., 1992; Cramez and Jackson, 2000; Marton et al., 2000; Valle et al., 2001). A further reactivation occurred in the late Miocene and Pliocene (Spathopoulos, 1996; Nzé Abeigne, 1997; Marton et al., 2000). Two main phases of compression are documented along the Angola Escarpment: Late Albian to Turonian, followed by an Oligocene to Present growth of the frontal bulge (Anka, 2004).

Evolution of the sedimentation rate through time suggests a close correlation with gravitational tectonics, on both margins, in spite of different tectonic styles **(Fig. 6)**. On the SW Africa margin, the long-lasting deformation that activated successively younger stratigraphic decollement surfaces, was driven by a slightly increasing sedimentation rate (Rust and Summerfield, 1990). In contrast, the extremely variable sedimentation rates on the equatorial western Africa margin (Leturmy et al., 2003), correlate with the Late Cretaceous and Oligocene-Present phases of salt tectonic. This is a clear indication of the dominant part played by sedimentary load on activation of gravitational tectonics (*e.g.* (Raillard et al., 1997). The other important parameter is the slope of the decollement, that is primarily controlled by tilting of the margin induced by thermally-driven subsidence and sediment loading, and modified by regional tectonics (Nzé Abeigne, 1997).

**5.2. Regional tectonics**

As seen above, post-rift deformation on both segments is dominated by gravitational tectonics. However, it is triggered by basement differential subsidence. Indeed, the rate of the thermally-driven, post-rift subsidence of the margin increases with the amount of synrift lithospheric stretching: therefore, subsidence at the landward end of the margin is slower than around the hinge zone, and results in westward tilting of the basement. Tilting of the evaporite decollement in the equatorial western Africa triggered early salt tectonics. The salt continued to move as long as high rate of carbonate production ensured continuous differential loading (Albian-Turonian).

In addition, tectonics involving deformation of the basement, driven by far-field stress changes at plate boundaries, or driven by the underlying asthenosphere can be identified in the stratigraphic record of the South Atlantic margins of Africa. On seismic reflection data, basement-involved deformation of the post-rift interval indicates regional tectonic movement **(Fig. 7)**. Previous studies on the south Gabon margin (Nzé Abeigne, 1997) has shown that transfer faults were generated during rifting and reactivated during Santonian, late Maastrichtian, and late Eocene. Similarly, the onshore Kwanza province display basement involved shortening structures interpreted as result of Senonian regional tectonics (Hudec and Jackson, 2002). Such deformations correlate with specific phases recorded in continental Africa (Guiraud et al., 1992), especially the Santonian event. Nzé Abeigne (1997) interpreted the latter as a result of the drastic drop in the rate of spreading at the mid-Atlantic ridge at magnetic anomaly 34 (Nürnberg and Müller, 1991) that induced a change in the intra-plate stress level (ridge push). According to (Guiraud and Bothworth, 1997), the Santonian unconformity, that is identified across the whole African plate, was generated by an inversion related to the change in movement for the opening of the North Atlantic.

A number of studies have argued the existence of several phases of post-rift uplift along the South Atlantic margin of Africa. Characterisation of uplift and erosion (amplitude, distribution and timing) is crucial in understanding the terrigenous sedimentation on the margin (Leturmy et al., 2003; Lucazeau et al., 2003). Following the rift-related Neocomian uplift, thermochronology (apatite fission track and fluid inclusions) give evidence for two main periods of post-rift uplift that affected the margin east of the hinge line: 1) during Senonian (Santonian and Maatrichtian), and 2) during Eocene to Late Miocene interval (Walgenwitz et al., 1990; Walgenwitz et al., 1992; Gallagher and Brown, 1999). The Santonian event documented by unconformities in the stratigraphic record of the margin and influx of terrigenous sediments (Teisserenc and Villemin, 1989) (**Fig. 7**), correlates with Santonian kilometre-amplitude denudation (interpreted as an uplift) of the coastal basin in Gabon (Walgenwitz et al., 1992). Further south, restoration of salt across the Kwanza Basin suggests Senonian basement uplift (Hudec and Jackson, 2004). These authors place the basement deformation around Campanian time. This can be linked with the modelled denudation of up to 3 km, that affected Namibia and South Africa, during the Late Cretaceous (Gallagher and Brown, 1999). Denudation occurring several hundreds of km inland, well beyond the boundary of the rifted lithosphere, as well as its timing (40-60my later than break-up) suggest a rifting-independent cause. The largest amplitude of erosion in SW Africa and the crystalline nature of the eroded basement may account for the much larger volume of Senonian sediments deposited on the SW Africa margin compared to the underfilled margin of equatorial western Africa (**Fig 6**), during the same interval. A subsidiary peak of sedimentation is recorded during Senonian; it correspond to a depocenter in Gabon, localised off a palaeo-Ogooue (Teisserenc and Villemin, 1989).

Deformation of passive markers such as Cenomanian shorelines provides values of post-Cenomanian finite uplift through continental Africa (Sahagian, 1988). It gives evidence for up to 3km total uplift in the onshore continuation of the Walvis Ridge, together with a generalised >1km uplift of the whole southern Africa, but the respective contributions of Senonian and later uplifts may be discussed. (Burke et al., 2003) argue that generalised uplift of Africa was initiated by the eruption of the Afar Plume at 31 Ma. According to these authors, the Oligocene submarine erosion followed by increased clastic sedimentation, observed on the South Atlantic margins of Africa, would be the result from this uplift. However, structural sections of the Kwanza Basin point to a Miocene uplift and erosion of some 1 to 2 km (Hudec and Jackson, 2004; Lunde et al., 1992). Flexural two-dimensional restoration of the Congo and northern Angola margins reveals Miocene uplift of the inner margin, with a maximum uplift rate occurring during Burdigalian and a secondary event in Tortonian (Lavier et al., 2000; Lavier et al., 2001). Thermochronology (fluid inclusions and AFT analysis) confirms the existence of Miocene uplift in the coastal basin of Gabon and Angola (Walgenwitz et al., 1990; Walgenwitz et al., 1992). However, unpublished AFT data acquired across continental southern Gabon show that outcropping basement rocks have cooled through the annealing zone (110°C to 60°C) during rifting, at the latest. Consequently, later denudation (including possible Tertiary event) did not exceed 1 to 1.5 km.

Although precise timing of the Tertiary uplift is less well constrained in Namibia and South Africa, amplitude of the associated denudation exceeds 1.5km and reaches 3km in extensive areas of Namibia (Gallagher and Brown, 1999). Correlation of Tertiary uplift distribution, and its decreasing amplitude away from Namibia, on one hand, with the large wavelength topography / bathymetry, geoid and high heat flow anomaly (Nyblade and Robinson, 1994; Hartley et al., 1996) on the other hand, strongly suggests that it is driven by asthenospheric movements, leading to the present-day superswell beneath SW Africa. If this causal relationship is accepted, stratigraphically constrained observations on the margins would indicate that the African superwell was emplaced in early Miocene, probably with additional increments during late Miocene.

## 6. Discussion

Some aspects of the different post-rift evolution of the two reviewed segments are still debated.

In particular the contrasting evolution of accumulation rate of terrigenous sediment in both segments is notable **(Fig. 8)**. Continued important accumulation rate on SW Africa margin during greenhouse period (Late Cretaceous-Eocene), compares with the low terrigenous flux to the equatorial western Africa, while the icehouse period (Oligocene - Present) sees rapidly increasing sedimentation in the equatorial western Africa (Leturmy et al., 2003) in contrast to the declining sedimentation rate in SW Africa margin (Rust and Summerfield, 1990). Although climate played a determinant part in this evolution, interactions with geodynamics and surface process (Lucazeau et al., 2003) have to be considered in order to understand the different response to climate change for the two neighbouring margin segments.

Topography of SW Africa margin is dominated by the Great Escarpment (1000 –1500 m altitude) that represents the inherited rift-shoulder uplift (Summerfield, 1991; Kooi and Beaumomt, 1994). Numerical models of coupled tectonic and surface processes (Gilchrist et al., 1994) indicate that 1) denudation mostly affects the coastal, west-facing, narrow catchment; 2) delayed denudation affects the much wider interior catchment once the rift shoulder is breached; 3) the bulk of denudation occurred before the Tertiary. Apatite fission track results support denudation of the western part of the Great Escarpment before 50Ma (Brown et al., 1990; Gallagher and Brown, 1999) by coastal rivers with regressive erosion of the rift shoulder. It thus accounts for the large siliciclastic late Cretaceous-Eocene depocentre **(Fig. 8)**.

Non-volcanic rifting in the equatorial western Africa affected a wide zone of the continental lithosphere **(Fig. 2)**, and at least two zones of lithospheric flexure ("hinge zones") have been documented (Karner and Driscoll, 1999). Vertical movement (basinward subsidence and landward uplift) was therefore distributed across the > 200km wide margin, and resulted in a rift shoulder of moderate amplitude. Consequently, moderate topography did not generate much erosion and terrigenous flux to the margin thus remained low during the late Cretaceous-Eocene interval.

Early Oligocene greenhouse-icehouse transition induced the change in continental erosion expressed by an increase in terrigenous sediment flux to the equatorial western Africa margin (Séranne, 1999) **(Fig. 8)**. It is argued that during icehouse conditions, high frequency alternate glacial / interglacial periods expressed by alternate wet / wetter climates is the major parameter in bedrock erosion (Knox, 1972). Analyses of alterites in Equatorial Africa indicate long term evolution from tropical climate during the late Cretaceous and Paleocene, toward more humid equatorial climate during Neogene (Tardy and Roquin, 1998), culminating in the Congo catchment during the Quaternary, with annual precipitation of 1600 mm. Such elevated precipitation rate induces high runoff values (> 40 litres per seconds) that favours mechanical and chemical weathering (Summerfield and Hulton, 1994). Increasing continental erosion throughout the Oligocene to Present is independently documented by the Strontium isotopic ratio: it increases when continental radiogenic granitoids are eroded and transferred to the oceans **(Fig. 8)**. The reduced sediment yield on the SW Africa margin, however, seems to contradict this model. In contrast to equatorial Africa, continental deposits in Namibia and South Africa suggest an evolution towards more arid climate during the Tertiary (Tardy and Roquin, 1998) (Scotese, 2002). In fact, very small values of denudation (0.5–1 m/My) during Neogene were due to the extreme aridity of the Namibe desert, following the establishment of the cold Benguela Current in mid-Miocene time (Wateren and Dunai, 2001). It is suggested here, that in spite of the high topographical gradient and of the enlargement of the Orange catchment that breached the rift shoulder, the local extremely arid climate prevented erosion of the SW Africa margin, during Oligocene to Present.

A phase of Miocene tectonic uplift affected the landward side of both reviewed margins and was associated with accelerated subsidence of the basin. Better stratigraphic resolution on the equatorial western Africa margin placed this event around early Miocene (Burdigalian). It is a response to deep-seated asthenospheric movements located beneath southern Africa and the local expression of the establishment of the "African superswell". Uplift of continental SW Africa had an effect on the stratigraphic architecture: by-pass or erosion of the shelf, sediment preservation on the slope and progradation of clastic wedge fed by increased continental erosion. Interestingly, these effects on the stratigraphic architecture are similar to those resulting from the installation of icehouse conditions as from early Oligocene (see above). The regional (African superswell) and global (greenhous-icehouse transition) phenomena acted in the same way and their effects are blended in the observed stratigraphy. The two phenomena have usually been confused (e.g. Burke et al., 2003) however, they can be distinguished by their relative chronology. In turn, the Tertiary reactivation of salt tectonics on the equatorial western Africa margin may be explained either by the additional tilting of the decollement resulting from the *circa* 500m uplift of the margin, east of the hinge-line (*e.g.* Walgenwitz et al., 1992), either by the increase of sedimentation that induced an increased sedimentary load (*e.g.* Duval et al., 1992). Most likely, the multiple reactivations of gravitational tectonics during the Oligocene to Present are a response of the interplay of increased continental erosion driven by climate change and associated increased sedimentary load (Séranne, 1999) and of tectonic uplift (Lavier et al., 2001).

Sedimentary loads, such as margins depocentres, applied on lithosphere with finite flexural rigidity, result in increased subsidence beneath the depocentre and flexural uplift at the periphery of the load (Watts, 1989). In spite of the limited amplitude of such flexural uplifts (tens of metres) in comparison to the load-induced subsidence (thousands of metres), it has important effects on the margin stratigraphy (Watts, 1989), and on the organisation of the river catchments feeding terrigenous sediments to the margin. For example, Whiting and others (1994) demonstrated the interaction of the load of the Indus fan on the Indian margin, and Driscoll and Karner (1994) investigated the modification of the coastal drainage around the Amazon fan. Modelling the flexural response of the Tertiary loads on the equatorial western Africa margin provides interesting results (Anka, 2004; Lucazeau et al., 2003). Application of the Congo deep-sea fan load in a distal position with respect to the margin, increases the subsidence of the offshore margin, and induces a flexural uplift up to 100 m of a strip parallel to the coast, several hundreds kilometres inland. In turn, the flexural uplift may favour erosion of the coastal catchments and contributes to sedimentation to the margin. Modelling with variable geologically reasonable parameters shows that this process alone cannot account for the observed margin's stratigraphy and onshore coastal denudation (Lucazeau et al., 2003; Anka, 2004). Interestingly, the flexural response of the lithosphere to Tertiary loads also contributes, along with the Oligocene climate change and the Burdigalian tectonic uplift, to the increased subsidence and sedimentation on the equatorial western Africa margin and uplift/erosion of the coastal river catchments.

## 7. Summary and conclusion

The present-day South Atlantic margins of Africa displays a striking symmetry of the two segments separated by the Walvis ridge: Two large interior basins (Congo and Orange River catchments) connected with thick sediment depocentres (Fig. 1). However, today's setting is misleading as it results from a contrasted evolution of the two segments **(Fig. 9)**. The marked difference of the South Atlantic margins of Africa, dates back to rifting (volcanic dominated rifting and high rift-shoulder vs wide zone of crustal streching). The distinct post-rift evolution of both segments was controlled by the rift-stage inherited structures. Differences in sedimentation arose from restriction of the equatorial western Africa by the Walvis ridge and from the existence of an important rift-shoulder in SW Africa margin. The onset of icehouse conditions in early Oligocene drastically modified both margins evolution, although expression of the climate change was opposed. Due to northward motion of Africa plate, the Neogene climate zoning placed equatorial western Africa under tropical wet conditions. Consequently, the Congo catchment , that had been under arid conditions until then, responded with increased erosion and developed the huge Congo deep-sea fan. On the opposite, and in spite of an elevated topography due to the inherited rift-shoulder and successive phases of uplift, the very dry climate of southwest Africa did not allow important erosion, and terrigenous sedimentation dropped on SW Africa margin.

South Atlantic margins of Africa, extending along strike some 35° in latitude, provide demonstrative case studies of varying responses to the interplay of climate/sedimentation/ tectonics. In order to address these topics it is necessary to widen the extension of continental margins, usually studied from coastal basin to continent-ocean boundary. Oceanic basins and distal deep-sea fans need to be included in continental margin studies since they contain a significant part of the stratigraphic record. Equally, river catchments of the adjacent continent belong to the continental margin: they are subjected to uplift and climate change that affects the sedimentary flux to the margin. Precise evaluation and quantification of these processes represent the next challenge for Earth scientists.

**Acknowledgements**

This paper summarises ideas that have emerged during many discussions with Michel Lopez and Francis Lucazeau. It also greatly benefited from the reviews of René Guiraud, Ian Davison and Patrick Unternehr. We thank Total and Ifremer for their kind support at various stages of the study. Finally, we are deeply indebted to the Guest Editors, especially René Guiraud, for continuous encouragement and an incredible amount of patience with late contributors.

**Figures captions**

**Fig. 1:** Basin segmentation along the South Atlantic margins of Africa. Margin depocentres coresspondding to the total (syn- and postrift) sedimentary cover are compiled from seismic surveys (Anka, 2004; Emery et al., 1975a; Emery et al., 1975b; Moulin, 2003) and isopachs are given in seconds TWT rather than in kilometres in order to avoid discrepancies resulting from very different seismic velocities. Onshore continental Africa drainage supplying terrigenous sediments to the margin are divided into small coastal river catchments (white dotted) and the two large Congo and Orange catchments (grey dotted). This map illustrates the respective volumetric importance of each segment, and the relationship between river catchments and some of the depocentres. Sections of Fig. 2 , 4  and 7 are located.

**Fig. 2:** Comparative crustal-scale cross sections of the two representative segments of the South Atlantic margins of Africa, based on seismic data from (Bauer et al., 2000; Moulin, 2003). This display the major difference between a non volcanic margin (a, Angola) representative of the equatorial western Africa margin, and a volcanic margin (b, Namibia), typical of the SW Africa margin.

**Fig. 3:** Sketch of the structural and stratigraphic relationship of the synrift succession in the equatorial western Africa margin (not to scale). Hinge zone is the coast-parallel, 10-20km wide band, across which an increased gradient of vertical movement is recorded. Landward of the "Atlantic hinge zone", were structures and stratigraphy are rather well constrained by borehole data, synrift packages consists of tilted fault-blocks displaying 2 diverging sequences of growth structures, covered by a basinward diverging, late-rift sequence. West of the hinge line, were basement rapidly plunges, the synrift succession displays sub-parallel reflectors, onlapping the basement. This sequence could either be a time equivalent to the 3 synrift sequences, or either correlate with the late-rift sequence.

**Fig. 4:** Generalised depth sections of the drift sequence across the equatorial western  Africa margin (a) and the SW Africa margin (b), displaying stratigraphy and gravitational tectonic structures. Sections are localised on Fig. 1.

**Fig. 5:** Isopach map of late-drift, post-Oligocene to Present sequence along the South Atlantic margins of Africa, compiled from (Anka, 2004; Rust and Summerfield, 1990), in relation with river catchments (as in figure 1). This map illustrates the diverging late-drift evolution of the two representative segments. North of the Walvis ridge, the respective size of the depocentres (mostly siliciclastic) mirror the extent of the river catchments, and the drastic increase in terrigenous sedimentation corresponds to the increased continental erosion rate from the Early Oligocene greenhouse-icehouse transition. South of the Walvis ridge, the much smaller post-Oligocene depocentre is not clearly related to individual river catchments, and the Oligocene climate change was expressed by a decrease in terrigenous flux to the margin.

**Fig. 6:** Correlation chart of gravitational tectonics in the SW Africa and equatorial western Africa margins, in relation with the sedimentation rate. In SW Africa margin, three decollements in shale levels allow extension (dotted interval) of the upper slope and correlative thrusting (light grey interval) in the basal slope. In the equatorial western Africa, there is one single decollement in the Aptian salt, that was activated during two main phases. The contrasted evolution of sedimentation rate (dark grey histogram) on the SW Africa margin (Rust and Summerfield, 1990) and the equatorial western Africa margin (Leturmy et al., 2003) correlates with the timing of gravitational tectonics, which suggests that sedimentary load acts as a primary control on gravitational tectonics. Time scale of (Gradstein and Ogg, 2004).

**Fig. 7:** Line drawing of a margin-parallel seismic profile across the N'Komi transfer zone in south Gabon margin. The diverging reflectors, different thickness of contemporaneous sequences, and onlaps on unconformities provides evidence for post-rift tectonic reactivation of faults bounding a basement high. Numbers refers to dated markers that are regionally mapped (Nzé Abeigne, 1997). Seismic stratigraphy indicates tectonic reactivation during Santonian (erosional unconformity 11) and other secondary phases of reactivation during Late Masstrichtian and Eocene.

**Fig. 8:** Correlation chart of accumulation rates, tectonics and proxies of global climate and erosion, in order to highlight possible causal effects and interactions between local / regional tectonics and global

forcing discussed in the text. Time scale of (Gradstein and Ogg, 2004). Dominant sedimentation type is indicated along with the accumulation rate (Anka, 2004; Leturmy et al., 2003; Rust and Summerfield, 1990). The width of individual tectonic event is a tentative estimate of the temporal evolution of its intensity; e.g.: the rate of thermally driven post-rift subsidence exponentially decreases with time. Oxygene isotopes (Zachos et al., 2001) indicates ocean temperature and the onset of icehouse conditions in early Oligocene. Formation of polar icecaps induced long-term sea-level drop level (Haq et al., 1988) that favorises progradation of clastic wedges. Strontium isotopic ratio in the oceans (Bralower et al., 1997; Elderfield, 1986) is a good indication of continental erosion, as a result of icehouse climatic conditions (Séranne, 1999). This chart shows that the early history of both margins is controlled by local tectonics, while the onset of icehouse conditions in early Oligocene drastically modified both margins evolution. The late-drift diverging evolution is discussed in the text.

**Fig. 9:** Cartoon of the comparative evolution of the erosion-sedimentation systems on the two segments of the South Atlantic margins of Africa, separated by the Walvis Ridge (not to scale). During Albian-Turonian and continuing through late Cretaceous, active erosion and clastic sedimentation on the SW Africa margin was controlled by the presence of a rift shoulder while low-lying hinterland of equatorial western Africa prevents active erosion and favours formation of a carbonate shelf. Following the early Oligocene greenhouse-icehouse transition (Séranne, 1999) and displacement of climate zoning due to northward motion of Africa (Scotese, 2002), the erosion-sedimentation systems are drastically modified. Wet climate in equatorial Africa promotes intense erosion and formation the huge Congo deep-sea fan, while extremely arid climate in SW Africa reduces erosion and the size of the corresponding depocentre.

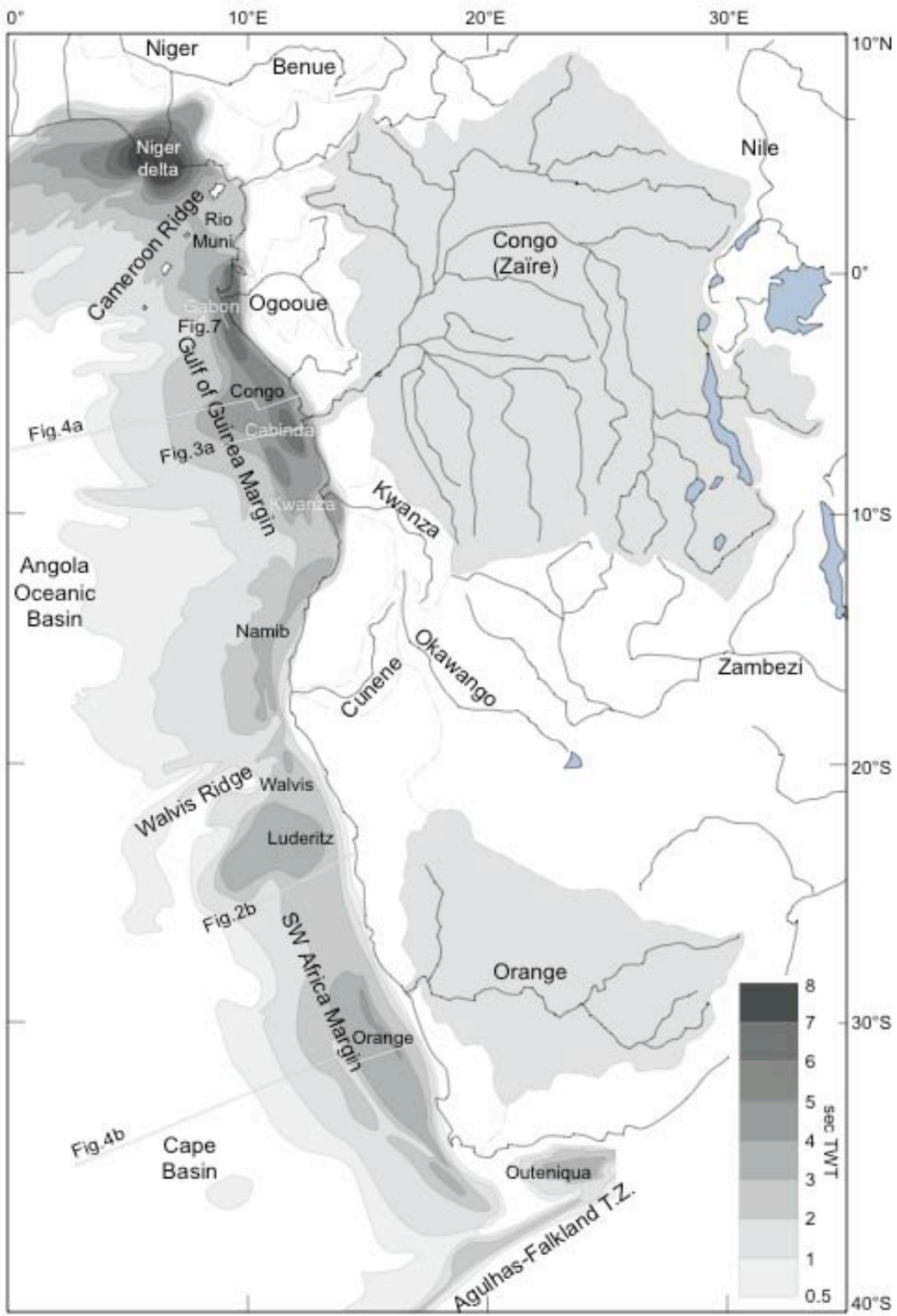

Séranne & Anka - Fig.1

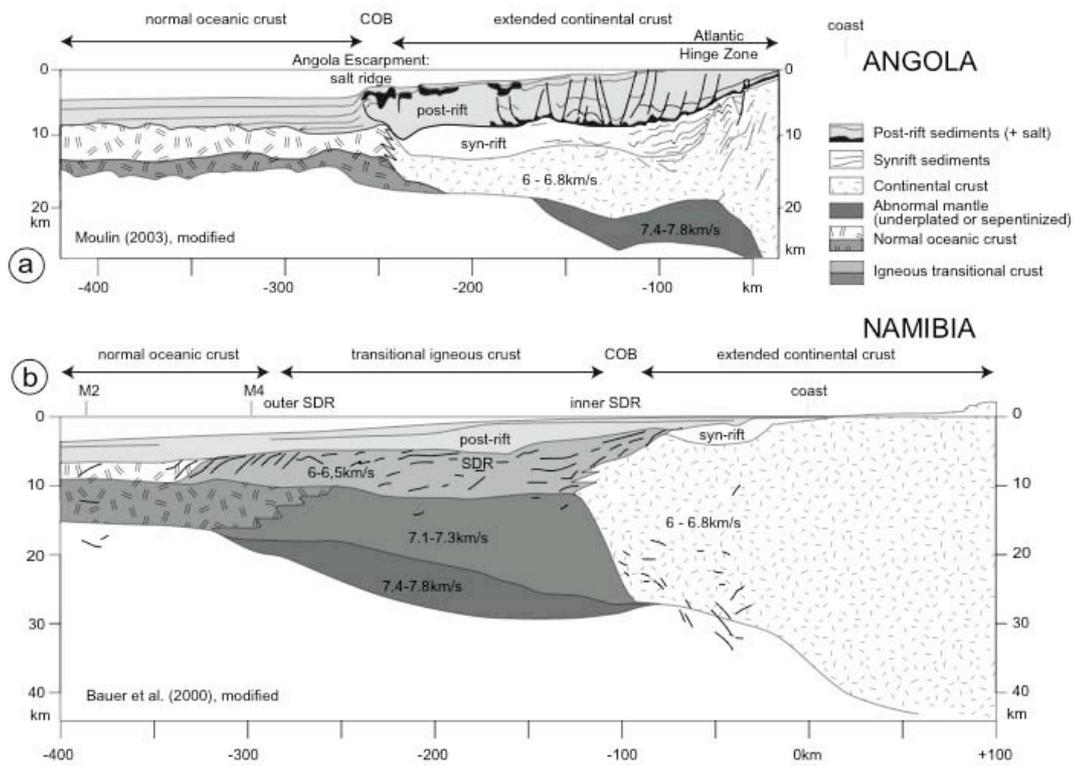

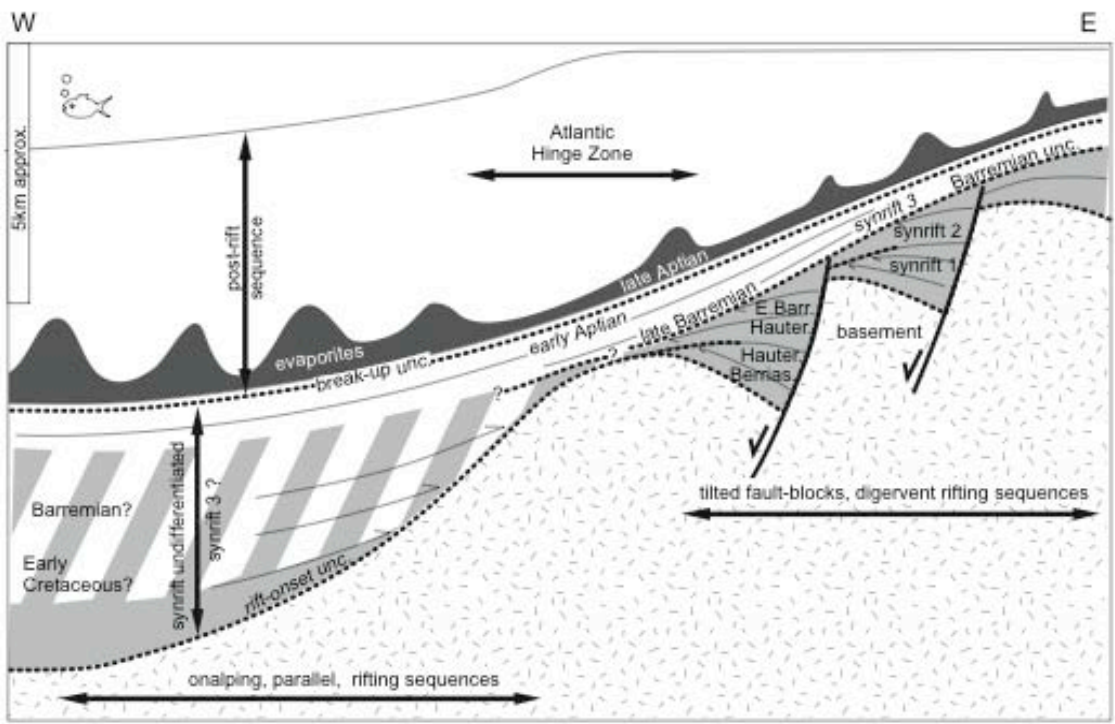

Séranne & Anka - Fig.3

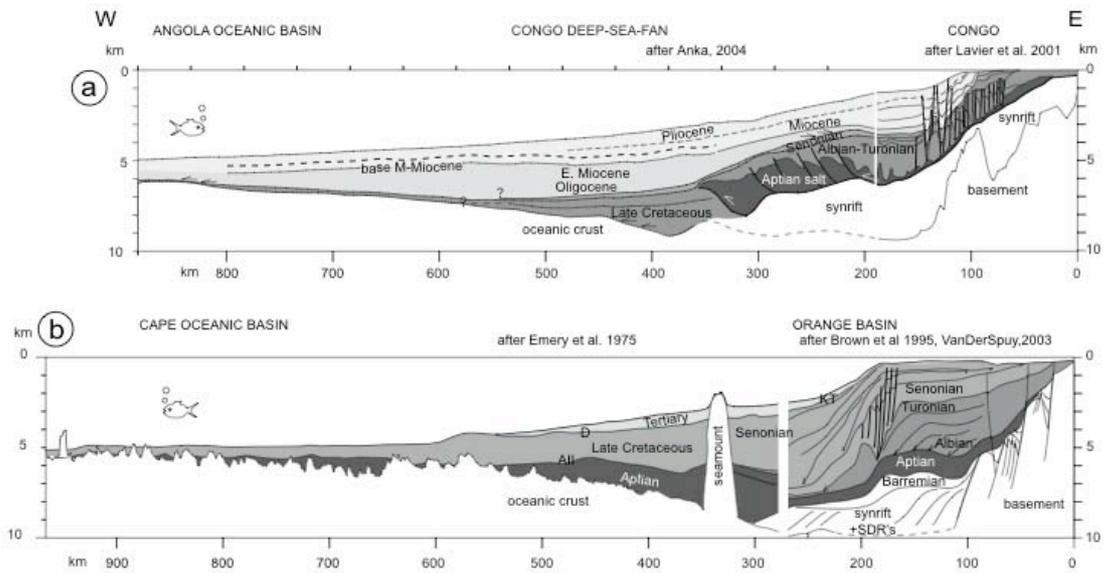

South Atlantic margins of Africa.	

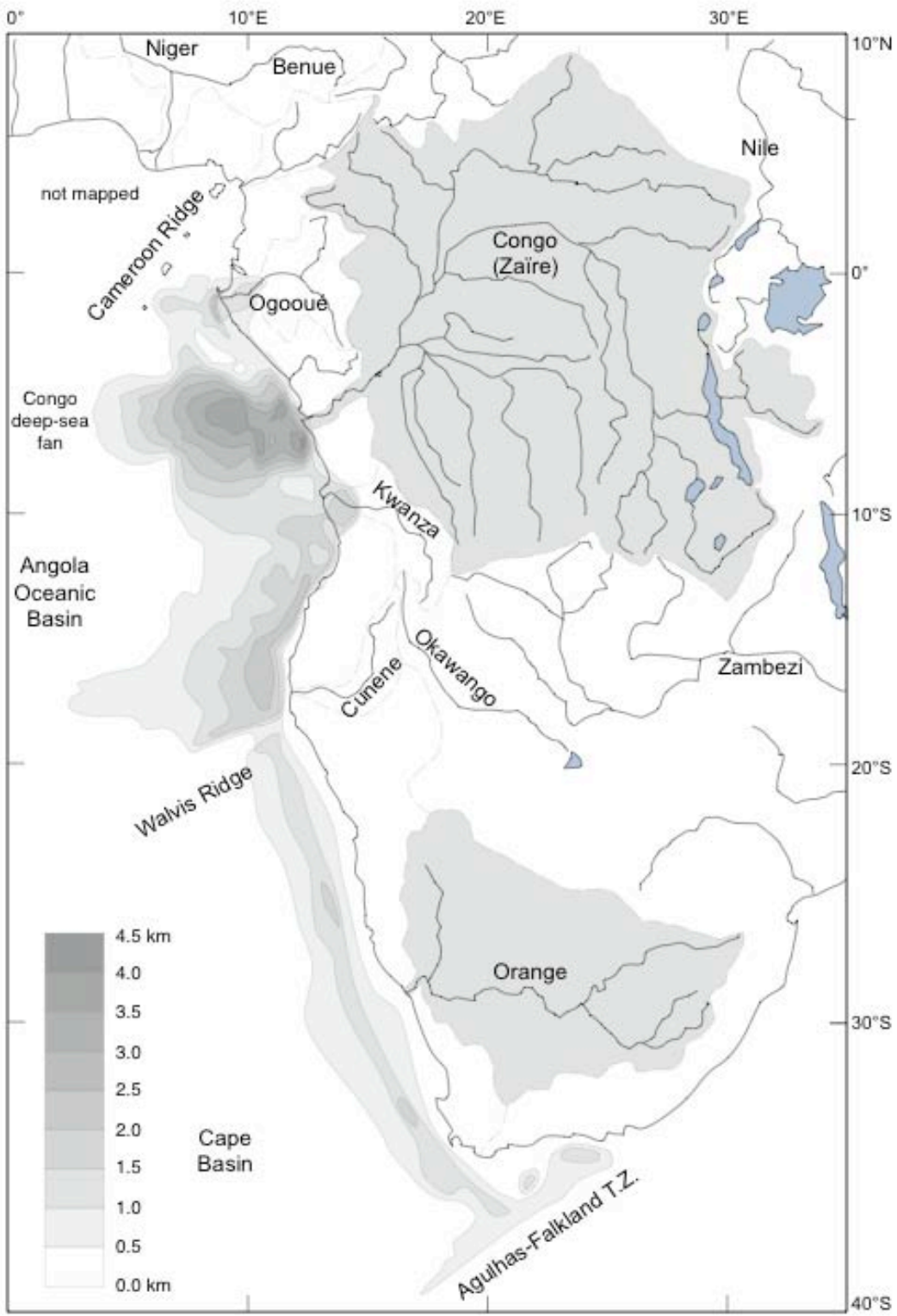

Séranne & Anka - Fig.5

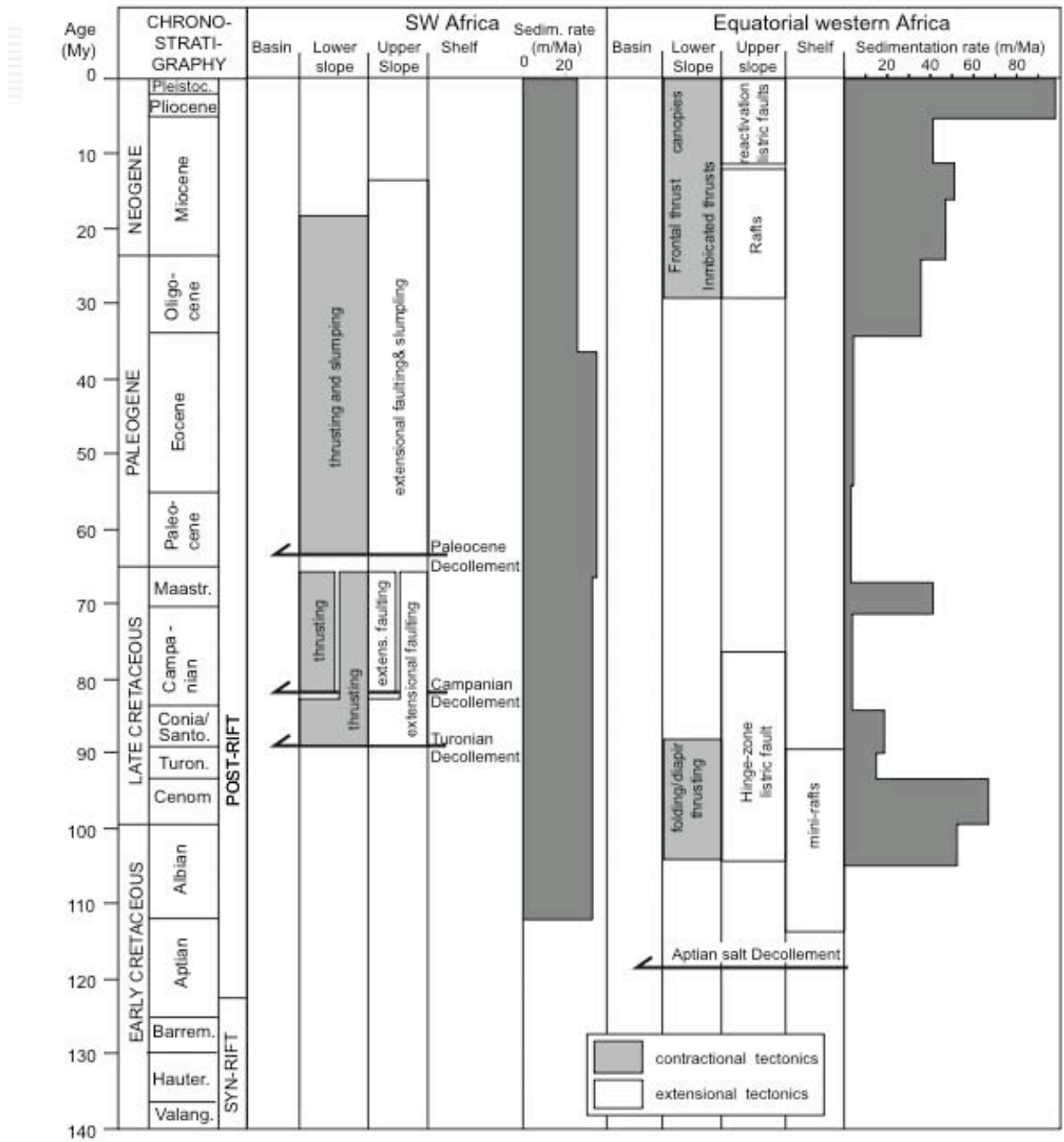

Séranne & Anka - Fig.6

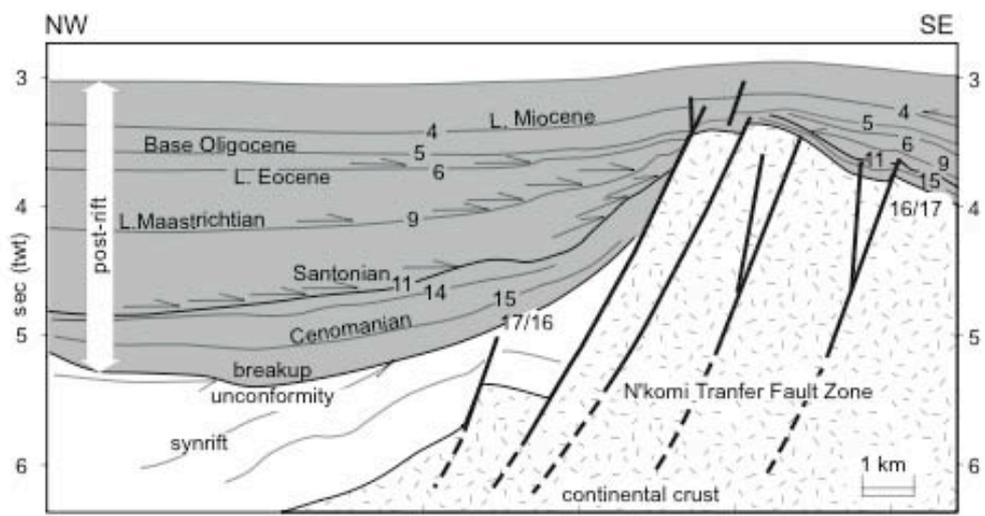

Séranne & Anka - Fig.7

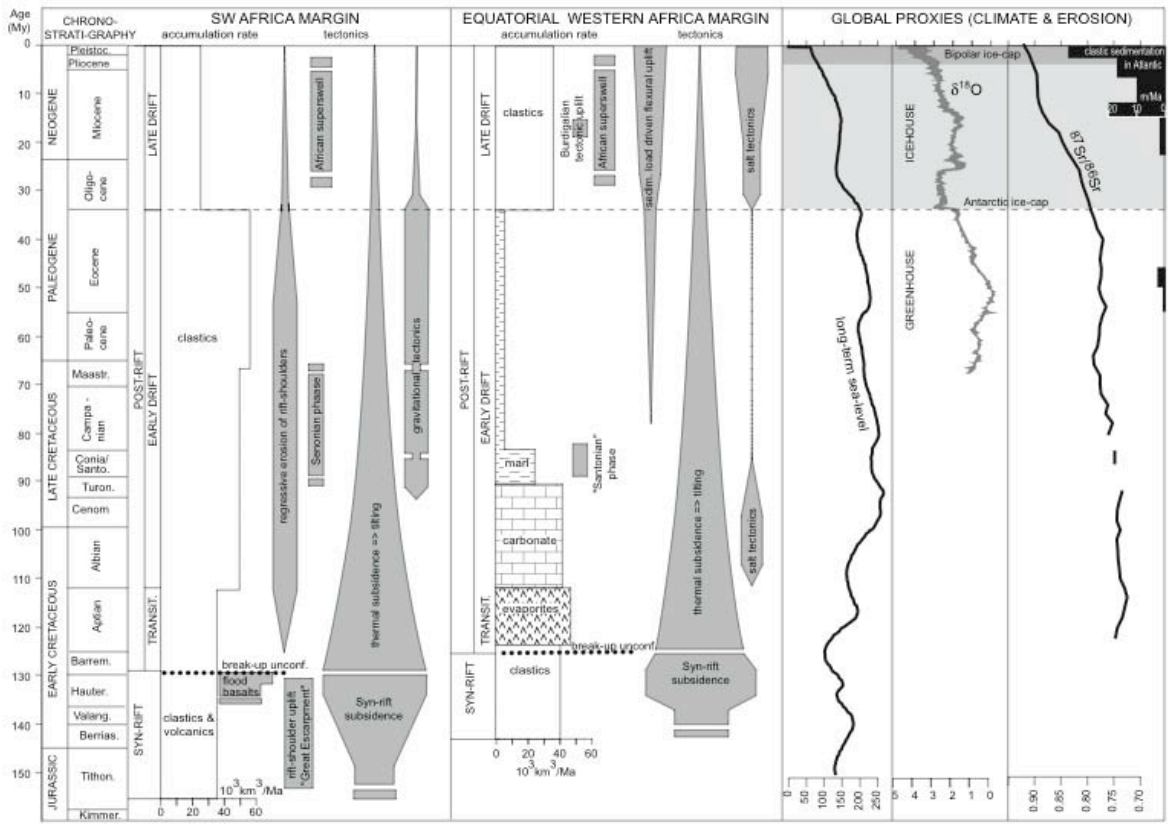

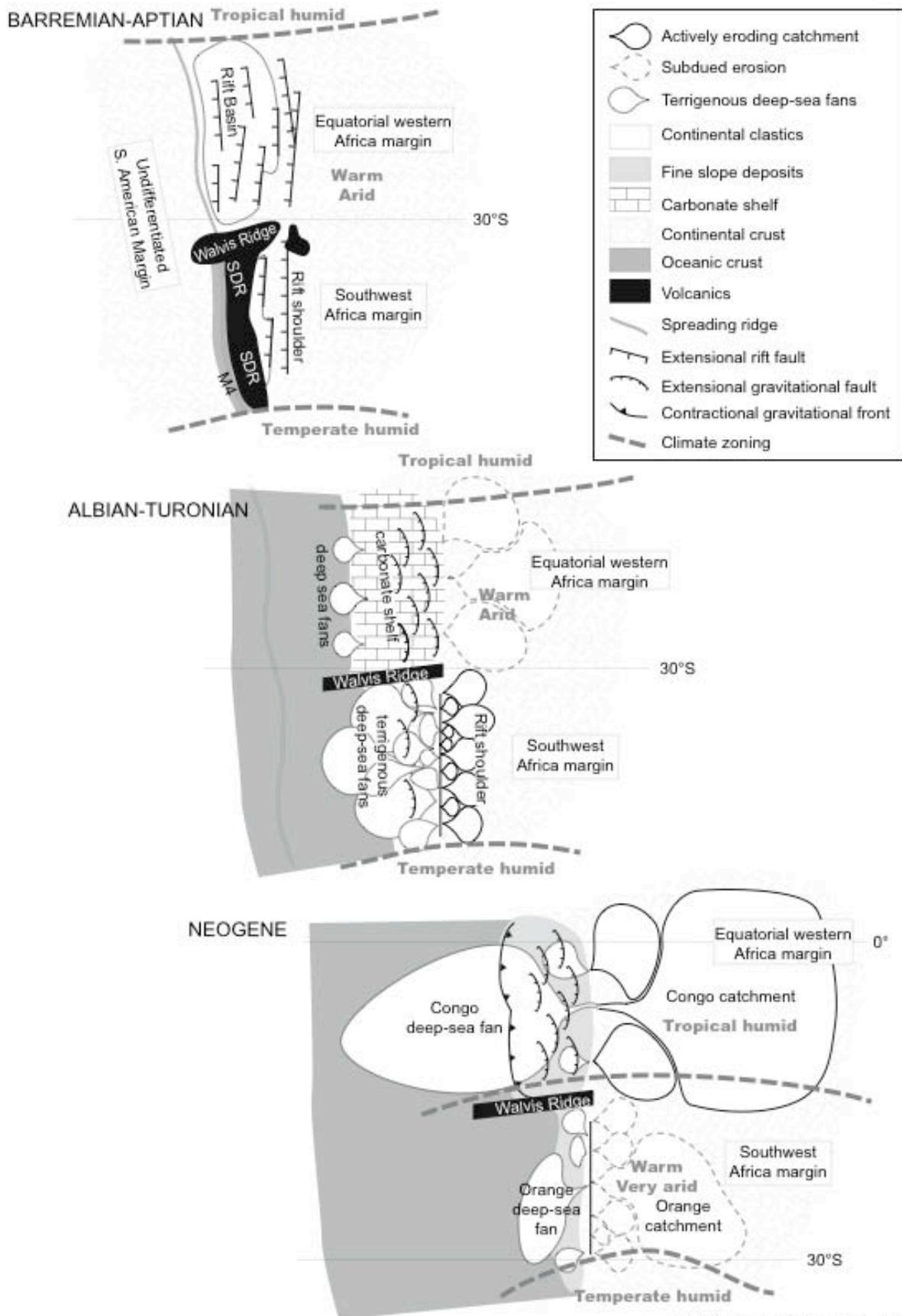

Séranne & Anka - Fig.9